\documentstyle[12pt,aasms4]{article}

\def\arcsecpoint{$''\!.$}

\received{1998 December 23}
\accepted{1999 March 29}
\lefthead{Crenshaw \& Kraemer}
\righthead{Intrinsic Absorption in NGC 5548}

\begin{document}

\title{Intrinsic Absorption Lines in the Seyfert 1 Galaxy NGC 5548:\\
UV Echelle Spectra from the Space Telescope Imaging
Spectrograph\altaffilmark{1}}

\author{D. Michael Crenshaw\altaffilmark{2,3}
and Steven B. Kraemer\altaffilmark{2,4,5}}

\altaffiltext{1}{Based on observations with the NASA/ESA {\it Hubble Space 
Telescope}, which is operated by the Association of Universities for Research in 
Astronomy, Inc., under NASA contract NAS5-26555.}

\altaffiltext{2}{Catholic University of America and
Laboratory for Astronomy and Solar Physics,
NASA's Goddard Space Flight Center, Code 681
Greenbelt, MD  20771.}

\altaffiltext{3}{Email: crenshaw@buckeye.gsfc.nasa.gov.}

\altaffiltext{4}{Email: stiskraemer@yancey.gsfc.nasa.gov.}

\altaffiltext{5}{Space Telescope Imaging Spectrograph (STIS) Instrument 
Definition Team}

\begin{abstract}

We present the first observations of a Seyfert galaxy with the echelle gratings 
on the Space Telescope Imaging Spectrograph (STIS), which provide
high-resolution ($\lambda$/$\Delta\lambda$ $\approx$ 40,000) coverage of the 
intrinsic UV absorption lines in NGC~5548. We confirm the presence of five 
kinematic components of absorption in L$\alpha$, C~IV, and N~V at radial 
velocities of $-$160 to $-$1060 km s$^{-1}$ with respect to the emission lines, 
and find an additional L$\alpha$ component near the systemic velocity, which 
probably arises in the interstellar medium of the host galaxy.
Compared to GHRS spectra of the N~V and C~IV absorption 
obtained $\sim$2 years earlier, the kinematic components have not changed in 
radial velocity, but the ionic column densities for two components have 
decreased. We attribute these variations to changes in the total column of gas, 
but for one component, we cannot rule out changes in the 
ionization of the gas.

We have calculated photoionization models to match the UV column densities from 
each of the five components associated with the nucleus. In four of the 
components, the ionization parameters (U $=$ 0.15 -- 
0.80) and effective hydrogen column densities (N$_{eff}$ $=$ 6.0 x 10$^{18}$ 
cm$^{-2}$ -- 2.8 x 10$^{20}$ cm$^{-2}$) cannot produce the O~VII and O~VIII 
absorption edges seen in the X-ray warm absorber. The remaining component 
is more highly ionized (U $=$ 2.4, N$_{eff}$ $=$ 6.5 x 10$^{21}$ cm$^{-2}$) and 
our model matches the previously observed X-ray absorption columns. This 
component is therefore likely to be responsible for the X-ray warm absorber.
It also has the highest outflow velocity and showed the largest variations in 
column density. 

\end{abstract}


\section{Introduction}

With the advent of the {\it Hubble Space Telescope} (HST), it has become clear 
that intrinsic absorption is a common phemonemon in the UV spectra of Seyfert 1 
galaxies. From a study of HST archive spectra, we determined that $\sim$60\% 
(10/17) of the Seyfert 1 galaxies in our sample have intrinsic absorption lines 
(Crenshaw et al. 1999). All ten of the Seyferts with absorption showed high 
ionization lines (N~V $\lambda\lambda$1238.8, 1242.8; C~IV 
$\lambda\lambda$1548.2, 1550.8), in addition to L$\alpha$. Low ionization lines 
were less common: four Seyferts showed detectable Si~IV $\lambda\lambda$1393.8, 
1402.8 absorption, and only one showed Mg~II absorption $\lambda\lambda$ 2796.3, 
2803.5 absorption (NGC 4151). The intrinsic absorption lines are blueshifted by 
up to 2100 km s$^{-1}$ with respect to the narrow emission lines, which 
indicates net radial outflow of the absorbing gas. At high spectral resolution, 
the lines often split into distinct narrow components, with widths in the range 
20 -- 400 km s$^{-1}$ (FWHM).

Intrinsic absorption is also common in the X-ray spectra of Seyfert galaxies; 
about half of these objects show ``warm absorbers'', characterized by O~VII and 
O~VIII absorption edges (Reynolds 1997; George et al. 1998) . We found a clear 
correspondence 
between the UV and X-ray absorption in our survey; of the eight Seyferts that 
were observed by both {\it HST} and {\it ASCA}, six showed both UV 
and X-ray absorption and two showed neither (Crenshaw et al. 1999).  Mathur and 
collaborators first established a connection between the UV and X-ray absorbers, 
and claimed that a single zone of photoionized gas can explain both the observed 
strengths of O VII and O VIII absorption edges and the UV absorption lines in 
quasars (Mathur 1994) and the Seyfert galaxy NGC 5548 
(Mathur et al. 1995). However, in NGC 4151 (Kriss et al. 1998) and NGC 
3516 (Kriss et al. 1996; Crenshaw et al. 1998), multiple zones spanning a wide 
range in ionization parameter and effective hydrogen column density are needed 
to explain the wide range in ionization species and large column densities of 
the UV absorption lines.

Since the intrinsic UV absorption in a Seyfert 1 galaxy is typically comprised 
of multiple kinematic components, an obvious question arises: are these 
components characterized by different physical conditions? An important related 
question is: to what extent do the UV and X-ray absorbers arise in the same 
regions?
In order to address these issues, near-simultaneous
high-resolution spectra of multiple lines are needed, to determine the column 
densities of different ions for each component. The only 
data that have met these requirements are the Goddard High Resolution 
Spectrograph (GHRS) observations of the C~IV and Mg~II 
absorption in NGC 4151 (Weymann et al. 1997; Kriss 1998). The C~IV/Mg~II column 
density ratio varies widely in this object, indicating a broad range in 
ionization parameter among the different kinematic components.

The Space Telescope Imaging Spectrograph (STIS) on {\it HST} offers an important 
means for investigating the differences in physical conditions among different 
kinematic components, by providing echelle gratings that 
cover broad bandpasses in the UV at a resolution of 
$\lambda$/$\Delta\lambda$ $\approx$ 40,000. To take advantage of this 
capability, we initiated a STIS Guaranteed Time Observations (GTO) program to 
obtain echelle spectra of several Seyfert 1 galaxies. Our first target is NGC 
5548.

\section{Observations and Analysis}

We obtained STIS echelle spectra of the nucleus of NGC 5548 on 1998 March 11 
through the 0\arcsecpoint2 x 0\arcsecpoint2 aperture. The observations are 
described in Table 1, along with previous GHRS observations of the N~V and C~IV 
regions obtained $\sim$2 years earlier (the N~V region contains a portion of the 
L$\alpha$ profile).
We reduced the STIS echelle spectra using the IDL software developed at NASA's 
Goddard Space Flight Center for the STIS Instrument Definition Team (Lindler et 
al. 1998). The procedures that we followed to identify the Galactic and 
intrinsic absorption lines and measure the intrinsic lines are given in Crenshaw 
et al. (1999).

Figure 1 shows the regions in the echelle spectra where the intrinsic absorption 
lines were detected (L$\alpha$, N~V, and C~IV ), and the regions where the 
strongest low-ionization lines might be expected (Si~IV, Mg~II). The fluxes are 
plotted as a function of the radial velocity (of the strongest member for the 
doublets) relative to the emission-line redshift, z $=$ 0.01676, obtained 
from the NASA/IPAC Extragalactic Database (NED). To obtain 
velocities relative to the redshift from H~I observations (z $=$ 0.01717), the 
velocity scale should be offset by an additional $-$123 km s$^{-1}$.
L$\alpha$ shows six distinct kinematic components, and the first 
five components are also seen in the absorption doublets of N~V and C~IV at 
essentially the same radial velocities.
The Si~IV and Mg~II regions show no obvious counterparts, but the spectra 
are important for putting upper limits on the column densities of 
these ions.

Components 1 -- 5 are the same as those identified by Crenshaw et 
al. (1999) in the GHRS spectra. Component 6 can also 
been seen in the GHRS spectra of L$\alpha$, but was not identified in that 
paper. Mathur, Elvis, and Wilkes (1999) have also identified the velocity 
components in the GHRS spectra of C~IV. Their identifications are the same as 
ours, except that they identify the dip in the red wing of component 4 as a 
separate component, and they identify the feature that we claim to be the C IV 
$\lambda$1550.8 line of component 5 with the C IV $\lambda$1548.2 line of a 
component that is slightly redshifted with respect to the systemic redshift.
The feature that corresponds to the $\lambda$1550.8 line of the redshifted 
component can be seen in the GHRS spectra (Crenshaw et al. 1999), but was too 
weak to satisfy our criteria for detection. We cannot identify this component in 
the STIS N~V and C~IV regions, but there is a weak feature at
$+$300 km s$^{-1}$ in the STIS L$\alpha$ region (Figure 1) that may correspond 
to Mathur et al.'s redshifted component. Due to the uncertainty about the 
existence of this component, we will not consider it further in this paper.

Comparing the STIS spectra with the GHRS spectra, we 
find that the absorption components are resolved in both.
The higher resolution of the STIS spectra makes the velocity structure in some 
of the components, such as the dip in the red wing of 
component 4, more distinguishable. 
We note that the short-wavelength GHRS spectrum does not include the blue wing 
of the L$\alpha$ and the absorption from component 1.
The GHRS spectrum of C~IV has a substantially higher SNR, since the exposure 
time was $\sim$3 times that of the STIS spectrum. Thus, the C~IV absorption for 
component 1 is not apparent in the STIS spectrum, and we can only give an 
upper limit on its column density (see below).

Table 2 gives the radial velocity centroids, widths (FWHM), and covering factors 
in the line of sight for each component. The values and uncertainties that we 
present are averages and standard 
deviations from individual lines; measurements of the individual lines and the 
Galactic lines will be given in another paper (Sahu et al. 1999).
The methods for determining the measurement errors are described in Crenshaw et 
al. (1999), and include uncertainties due to different reasonable placements of 
the underlying emission.
For each 
component, two lower limits are given for the 
covering factor in the line of sight: C$_{los}$, which is the fraction of total 
emission (continuum plus broad-line emission) that is occulted, and 
C$_{los}^{BLR}$, which is the fraction of broad-line emission that is occulted 
(assuming the entire continuum source is occulted). These lower limits have 
been determined from the residual intensities in the L$\alpha$ cores (see 
Crenshaw et al. 1999).

A comparison of the measurements in Table 2 with those from the GHRS spectra 
(Crenshaw et al. 1999) shows that, to within the errors, there have been no 
changes in the velocity centroids or widths of the components over $\sim$2 
years. The lower limits to the covering factors are essentially the same as 
those obtained from the GHRS spectra. For components 1 --5, both the total 
covering factor, C$_{los}$, and the BLR covering factor, C$_{los}^{BLR}$, are 
greater than one-half and in some cases, close to one. We 
can conclude that each of the components is likely to be outside of the BLR and 
comparable in size to or larger than the BLR in the plane of the sky. Component 
6 shows weak L$\alpha$ absorption, is not detected in the other 
lines, and is located close to the systemic velocity of the host galaxy, which 
suggests that the gas responsible for this component is associated with the 
interstellar medium of the host galaxy. We will not discuss this component 
further.

In Table 3, we give the column densities of L$\alpha$, N~V, and C~IV for each 
component. Since we only have lower limits to the covering factors, these values 
were determined by assuming that C$_{los}$ $=$ 1, and integrating the optical 
depths as a function of radial velocity across each component (see Crenshaw et 
al. 1999 for a discussion of the effects of C$_{los}$~$\neq$~1 on the measured
column densities).
The blending of the L$\alpha$ components in Figure 1 indicates that they are 
more saturated than the other lines, and therefore the column densities are 
affected by the deblending of the components and by errors in the removal 
of scattered light. The uncertainties in the L$\alpha$ columns in Table 3 are 
due to measurement errors, and do not include these effects.

Table 3 shows that there have been significant decreases in the N~V column 
densities for components 1 and 3, and a possible decrease in the C~IV column 
density for component 3, 
since the GHRS observations from $\sim$2 years earlier. These changes are also 
apparent in a comparison of Figure 1 with the GHRS spectra in Crenshaw et al. 
(1999). Since we only have an upper limit for the C IV column density of 
component 1, we have no information on its variability. The other components 
have not changed, given the uncertainties. 

To compare our results with those obtained for the X-ray warm absorber in NGC 
5548, we use the {\it ASCA} observations of this Seyfert on 1993 July 27 
(Reynolds 1997; George et al. 1998). From the optical depths of the O~VII and 
O~VIII absorption edges in Reynolds and the ionization cross sections, we 
calculate column densities of 1.0 x 10$^{18}$ and 1.6 x 10$^{18}$ cm$^{-2}$ for 
O~VII and O~VIII, respectively.

\section{Photoionization Models}
  
We have generated photoionization models to investigate the physical conditions 
in the absorption components. The details of the photoionization code are 
described in 
Kraemer et al. (1994, and references therein).  
Most of the input parameters for these models (e.g., solar abundances)
are identical to those used in our previous study of the narrow emission line
region (NLR) in NGC 5548 (Kraemer et al. 1998). We have modeled
the ionizing continuum as a broken power-law, $L_{\nu}$$=$K$\nu^{\alpha}$, 
with the following spectral indices:
$\alpha$ $=$ $-$1.0 (~h$\nu$ $<$ 13.6 eV),
$\alpha$ $=$ $-$1.4 (13.6 eV $\leq$ h$\nu$ $<$ 1300 eV), and
$\alpha$ $=$ $-$0.9 (~h$\nu$ $\geq$ 1300 eV).
Note that we have modified the value of $\alpha$ above 1.3 keV from that used in
Kraemer et al. (1998), based on Reynolds's (1997) fit to the 2 -- 10 keV 
continuum. 

The parameters that we varied in generating our photoionization models were
the ionization 
parameter U (the number of ionizing photons per hydrogen atom at the illuminated 
face of the cloud), and the effective hydrogen column density N$_{eff}$ (i.e., 
the neutral plus ionized hydrogen column). 
The ionic column densities predicted by the models do not depend on the atomic 
density, n$_{H}$.
For simplicity we chose a fixed value of n$_{H}$ $=$ 5 x 10$^{5}$ cm$^{-3}$; at 
the distance of our innermost component in the NLR models ($\sim$1 pc, see 
Kraemer et al. 1998), this yields a value of U $=$ 0.60 , which is 
approximately correct for producing the observed ratios of 
N~V to C~IV.
We generated a single model for each kinematic component, by varying
U until the N~V/C~IV column ratio was matched.  We then adjusted
N$_{eff}$ to fit the ionic column densities. We varied these parameters until 
the predicted columns matched the observed values to within the errors. Due to 
our concerns about the L$\alpha$ column densities (see the previous section), we 
did not use these columns to constrain the models.

The model values for components 2 -- 5 are listed in Table 4. Since we only have 
an upper limit to the C~IV column density for component 1 in the STIS data, we 
will treat that case separately. The predicted column densities for Si~IV and 
Mg~II are well below detectability in all cases; the largest value computed 
for Si~IV was $\sim$ 4 x 10$^{9}$cm$^{-2}$ (component 4), and the Mg~II column 
was effectively zero for all components.
Despite our concerns about the observed H~I columns, the observed and predicted 
values are reasonably close; the 
predicted values tend to be higher (by up to a factor of $\sim$2.4 for 
component 4).
One possible explanation is that our L$\alpha$ measurements are indeed affected  
by saturation effects; another is that the abundances are greater than solar by 
a factor of two
\footnote{Since this is the first time that we have run models in this 
high-ionization regime, we compared our calculations with those from CLOUDY90 
(Ferland et al. 1998).  The ionic column densities for the high ionization lines 
were quite similar, and the H~I columns from CLOUDY were a factor of $\leq$2 
higher than ours (due to a different treatment of the cooling), which indicates 
that our overprediction of the H~I columns is not a model artifact.}.
The values of U and N$_{eff}$ that we obtain for these components are well below 
those associated with typical X-ray warm absorbers (U $=$ 1 -- 10, N$_{eff}$ $=$ 
10$^{21}$ -- 10$^{23}$ cm$^{-2}$; Reynolds 1997; George et al. 1998).
To investigate this issue further, we computed
column densities for O~VII and O~VIII, which are also listed in Table 4.
The total predicted columns for these ions are 1.6 x 10$^{17}$cm$^{-2}$ and
9.9 x 10$^{15}$, whereas the observed values are $\sim$6 and $\sim$160 times 
higher for O~VII and O~VIII, respectively, which shows that components 2 -- 5 
do not contribute significantly to the X-ray warm absorber. 

For component 3, the N~V and C~IV column densities have decreased by about the 
same amount in the STIS data (factor of $\sim$1.6) compared to the GHRS 
observations (Table 2). This would indicate that the ionization parameter was 
essentially the same on these two occasions. Thus, the most likely explanation 
for the absorption changes is that the effective column density changed, due to 
bulk motion of some of the absorbing gas out of the line of sight. A possible 
discrepancy with this interpretation is that the H~I column density did not 
change appreciably, but we have already noted the difficulties involved in 
determining the H I columns from L$\alpha$.

For component 1, we generated a model based on the GHRS column densities. Table 
5 shows these values. Due to the large N~V/C~IV ratio, the 
ionization parameter and effective column density for this component are much 
higher than those for the other components. The predicted O~VII and O~VIII 
column densities are very close to the observed values, suggesting that this 
component is likely to be responsible for the X-ray warm absorber. Our values 
of U and N$_{eff}$ for this component are similar to those determined for the 
warm absorber in NGC 5548 by Reynolds (1997) and George et al. (1998). 
With the caveat that none of the UV or X-ray observations are simultaneous, we 
conclude that component 1 is likely to be the X-ray warm absorber.

To investigate the column density variations in component 1, we generated two 
additional models for the STIS data, as shown in Table 5. For the first model, 
we assumed that U did not change between the two observations, and for the 
second model, we assumed that N$_{eff}$ remained constant. Both models match the 
observed N~V column density, predict a low C~IV column, and overpredict the 
neutral hydrogen column.
Thus, the difference in ionic column densities
between the GHRS and STIS data for Component 1 could be simply explained by:
1) an increase of U by a factor of $\sim$ 1.4, or
2) a decrease of N$_{eff}$ by a factor of $\sim$ 4 along the line of sight. We 
prefer the latter explanation, since the discrepancy between the observed and 
predicted neutral hydrogen columns is similar to that for the other components, 
but cannot rule out the former.

\section{Conclusions}

We have obtained UV echelle spectra of NGC 5548 with STIS, and have confirmed 
the presence of five kinematic components of intrinsic absorption in the lines 
of L$\alpha$, C~IV, and N~V. An additional L$\alpha$ component is present near 
the systemic redshift, and is likely to be associated with the interstellar 
medium in the host galaxy. These components have not changed in their radial 
velocity coverage since the GHRS observations $\sim$2 years earlier.
The column densities of N~V and C~IV in component 3 have decreased, and the 
column density of N~V in component 1 has decreased over the two year interval 
(only an upper limit is available for component 1's C~IV column in the STIS 
data).

We have used photoionization models to examine the physical conditions in the 
gas responsible for the intrinsic absorption. The ionization parameters and 
effective column densities for components 2 -- 5 are lower than those associated 
with X-ray warm absorbers, and the predicted O~VII and O~VIII column densities 
for these components are too small to make a significant contribution to the 
X-ray absorption.
From the GHRS observations of N~V and C~IV, we obtain much higher values of U 
(2.4) and  N$_{eff}$ (6.5 x 10$^{21}$ cm$^{-2}$) for component 1. Our predicted 
O~VII and O~VIII columns for component 1 match the previous observed values. We 
conclude that this component is likely to be responsible for the X-ray warm 
absorber. 
This component also has the highest outflow velocity and exhibits the strongest 
variability. Since the GHRS C~IV and N~V spectra were obtained $\sim$6 months 
apart, and the {\it ASCA} observations were obtained $\sim$3 years earlier, 
these conclusions need to be checked with simultaneous UV and X-ray 
observations.

We find that the decrease in the N~V and C~IV column densities in component 3 is 
due to a change in the total column density of the absorption, rather than a 
change in ionization, since the N~V/C~IV ratio did not change significantly. 
This is the most likely explanation for the decrease in N~V column in component 
1 as well, although we cannot rule out a change in ionization. Assuming that 
bulk motion of the gas across the line of sight is responsible for the 
variations, complete coverage of the BLR, and a diameter of the N~V emitting 
region of $\sim$4 light days (Korista et al.), we find a transverse velocity of 
v$_{T}$ $\geq$ 1650 km s$^{-1}$. If the variations are due to bulk motion, these 
two components have transverse velocities which are comparable to their outflow 
velocities.

\acknowledgments

We thank Ian George for helpful discussions on the X-ray absorption.
D.M.C. and S.B.K. acknowledge support from NASA grant 
NAG 5-4103. Support for this work was provided by NASA through grant number 
NAG5-4103 and grant number AR-08011.01-96A from the Space Telescope Science 
Institute, which is operated by the Association of Universities for Research in 
Astronomy, Inc., under NASA contract NAS5-26555.

\clearpage

\figcaption[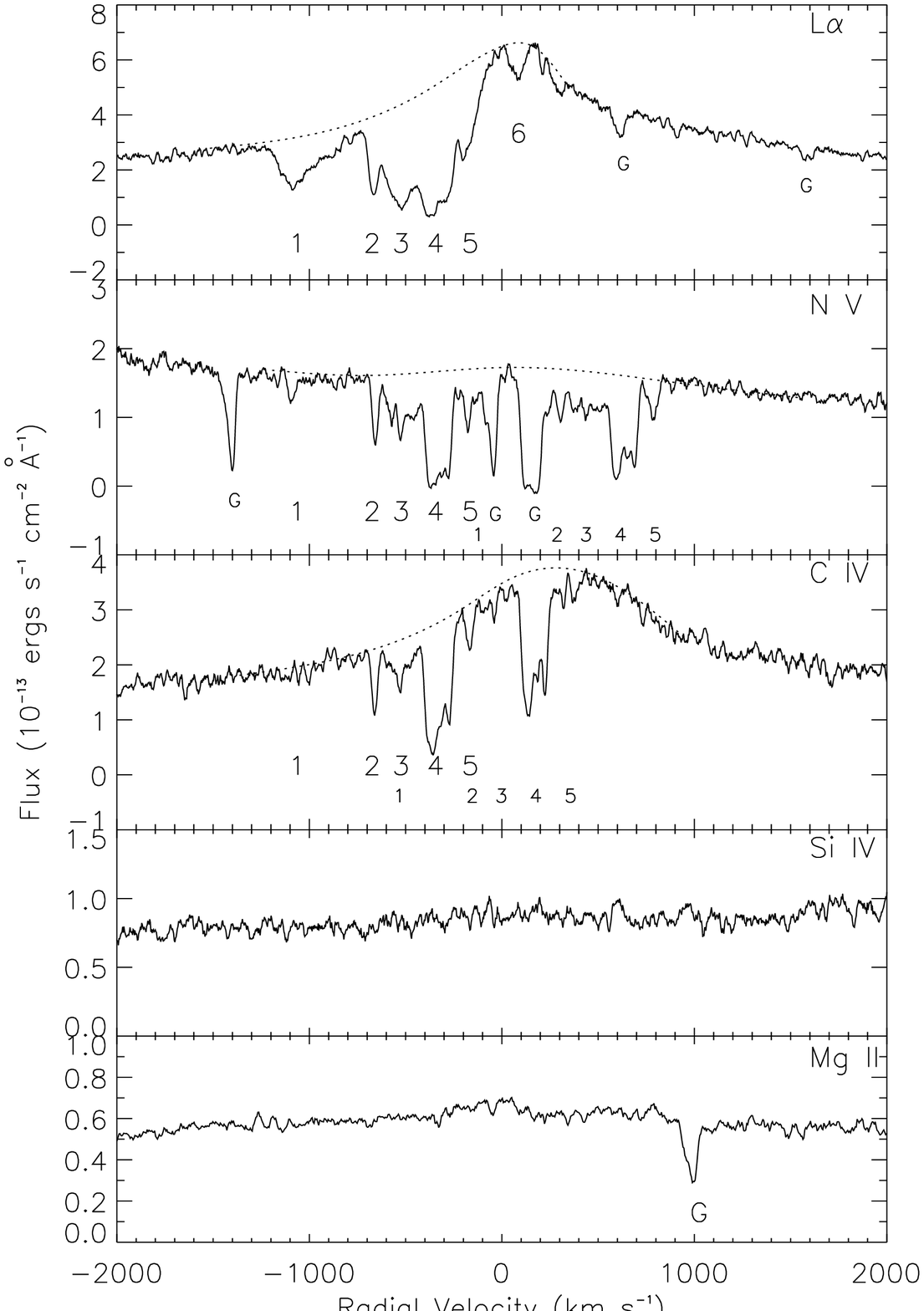]{Portions of the {\it STIS} UV echelle spectra of NGC 5548, 
showing the intrinsic absorption lines. Fluxes are plotted as a function of the 
radial velocity (of the strongest member, for the doublets), relative to an 
emission-line redshift of 0.01676. The kinematic components are identified with 
large numbers. Galactic lines are indicated with a ``G'', and the small numbers 
identify the weak members of the doublets. Fits to the emission profile are
given as dotted lines.}

\begin{deluxetable}{cccccl}
\tablecolumns{6}
\footnotesize
\tablecaption{High-Resolution UV Spectra of NGC 5548\label{tbl-1}}
\tablewidth{0pt}
\tablehead{
\colhead{Instrument} & \colhead{Grating} & \colhead{Coverage} &
\colhead{Resolution} & \colhead{Exposure} & \colhead{Date} \\
\colhead{} & \colhead{} & \colhead{(\AA)} &
\colhead{($\lambda$/$\Delta\lambda$)} & \colhead{(sec)} & \colhead{(UT)}
}
\startdata
STIS &E140M &1150 -- 1730 &46,000 &4750   &1998 March 11 \\
STIS &E230M &1607 -- 2366 &30,000 &2295   &1998 March 11 \\
STIS &E230M &2274 -- 3119 &30,000 &1905   &1998 March 11 \\
GHRS &G160M &1232 -- 1269$^{a}$ &20,000 &4607   &1996 February 17 \\
GHRS &G160M &1554 -- 1590$^{b}$ &20,000 &13,600 &1996 August 24
\tablenotetext{a}{N~V and a portion of L$\alpha$, Savage et al. 1997; Crenshaw 
et al. 1999}
\tablenotetext{b}{C~IV, Crenshaw et al. 1999; Mathur et al. 1999}
\enddata
\end{deluxetable}

\begin{deluxetable}{ccccc}
\tablecolumns{5}
\footnotesize
\tablecaption{Kinematic components and covering factors\label{tbl-2}}
\tablewidth{0pt}
\tablehead{
\colhead{Component} & \colhead{v$_{r}$$^{a}$} & \colhead{FHWM} &
\colhead{C$_{los}$} & \colhead{C$_{los}^{BLR}$}  \\
\colhead{} & \colhead{(km s$^{-1}$)} &
\colhead{(km s$^{-1}$)} & \colhead{} & \colhead{}
}
\startdata
1 &$-$1056$\pm$27 &85$\pm$25 &$>$0.62 &$>$0.51 \\
2 &$-$669$\pm$6   &39$\pm$9  &$>$0.76 &$>$0.71 \\
3 &$-$540$\pm$16  &78$\pm$25 &$>$0.90 &$>$0.88 \\
4 &$-$340$\pm$5   &140$\pm$5 &$>$0.94 &$>$0.93 \\
5 &$-$163$\pm$25  &75$\pm$7  &$>$0.63 &$>$0.58 \\
6$^{b}$ &~~88$\pm$5 &63$\pm$8  &$>$0.23 &$>$0.14
\tablenotetext{a}{Relative to the emission-line redshift of 0.01676.}
\tablenotetext{b}{Present only in L$\alpha$.}
\enddata
\end{deluxetable}

\begin{deluxetable}{cccc}
\tablecolumns{4}
\footnotesize
\tablecaption{Column densities in NGC 5548\label{tbl-3}}
\tablewidth{0pt}
\tablehead{
\colhead{Ion} &\colhead{Component} &\colhead{N (STIS)} &
\colhead{N (GHRS)} \\
\colhead{} &\colhead{} &\colhead{(10$^{14}$ cm$^{-2}$)} &
\colhead{(10$^{14}$ cm$^{-2}$)} 
}
\startdata
L$\alpha$ &1 &1.34$\pm$0.16 & ---------$^a$\\
L$\alpha$ &2 &0.61$\pm$0.12 & 0.76$\pm$0.09\\
L$\alpha$ &3 &2.01$\pm$0.11 & 1.79$\pm$0.21\\
L$\alpha$ &4 &2.97$\pm$0.12 & 3.30$\pm$0.29\\
L$\alpha$ &5 &0.66$\pm$0.08 & 0.77$\pm$0.11\\
L$\alpha$ &6 &0.13$\pm$0.04 & 0.13$\pm$0.04\\
N~V  &1 &0.44$\pm$0.18 &1.98$\pm$0.29 \\
N~V  &2 &0.82$\pm$0.16 &0.57$\pm$0.17 \\
N~V  &3 &2.33$\pm$0.32 &3.93$\pm$0.27 \\
N~V  &4 &8.54$\pm$0.65 &6.47$\pm$0.50 \\
N~V  &5 &1.18$\pm$0.24 &1.05$\pm$0.23 \\
C~IV &1 &$<$0.17~~~~~~~~~~ &0.11$\pm$0.04 \\
C~IV &2 &0.34$\pm$0.07 &0.28$\pm$0.04 \\
C~IV &3 &0.30$\pm$0.10 &0.48$\pm$0.15 \\
C~IV &4 &2.95$\pm$0.24 &2.89$\pm$0.22 \\
C~IV &5 &0.31$\pm$0.13 &0.41$\pm$0.18 \\
Si~IV & &$<$0.10 \\
Mg II & &$<$0.02 \\
\tablenotetext{a}{Not included in wavelength coverage.}
\enddata
\end{deluxetable}

\clearpage
\begin{deluxetable}{cccccccc}
\tablecolumns{8}
\footnotesize
\tablecaption{Model Predictions for STIS Ionic Column Densities\label{tbl-4}}
\tablewidth{0pt}
\tablehead{
\colhead{Comp.} & \colhead{U} 
& \colhead{N$_{eff}$}
& \colhead{H~I}
& \colhead{C~IV}
& \colhead{N~V}
& \colhead{O~VII}
& \colhead{O~VIII} \\
\colhead{}
& \colhead{}
& \colhead{(cm$^{-2}$)}
& \multicolumn{5}{c}{(10$^{14}$ cm$^{-2}$)}
}
\startdata
2   & 0.15   & 6.0$\times10^{18}$ &  0.72 &  0.35 &  0.85 & 20.10
&  1.26\\
3   & 0.80   & 2.8$\times10^{20}$ &  3.77 &  0.32 &  2.53
& 1.16$\times10^{3}$ & 56.20\\
4   & 0.20   & 8.3$\times10^{19}$ &  7.05 &  2.97 &  8.46 
& 3.28$\times10^{2}$ & 28.23\\
5   & 0.30   & 2.2$\times10^{19}$ &  1.11 &  0.33 &  1.25 
& 1.15$\times10^{2}$ & 13.70\\
\enddata
\end{deluxetable}

\begin{deluxetable}{lccccccc}
\tablecolumns{8}
\footnotesize
\tablecaption{Model Predictions for Ionic Column Densities$^{a}$
for Component 1
\label{tbl-5}}
\tablewidth{0pt}
\tablehead{
\colhead{Source} & \colhead{U} 
& \colhead{N$_{eff}$}
& \colhead{H~I}
& \colhead{C~IV}
& \colhead{N~V}
& \colhead{O~VII}
& \colhead{O~VIII} \\
\colhead{}
& \colhead{}
& \colhead{(cm$^{-2}$)}
& \multicolumn{5}{c}{(10$^{14}$ cm$^{-2}$)}
}
\startdata
 GHRS  & 2.42  & 6.5$\times10^{21}$ & 14.60 & 0.11 & 2.02 & 1.05$\times10^{4}$
& 2.29$\times10^{4}$\\
 STIS (U fixed) & 2.42   & 1.7$\times10^{21}$  & 3.63  & 0.02  & 0.41
& 2.29$\times10^{3}$ & 5.90$\times10^{3}$\\ 
 STIS (N$_{eff}$ fixed) & 3.42  & 6.5$\times10^{21}$  & 7.77  & 0.02  & 0.40
& 2.30$\times10^{3}$ & 2.10$\times10^{4}$\\  
\enddata
\end{deluxetable}

\clearpage
\begin{figure}
\plotone{fig1.eps}
\\Fig.~1.
\end{figure}

\end{document}